\documentclass[prd,twocolumn]{revtex4}

  \usepackage[usenames,dvipsnames]{pstricks}
\usepackage{amsmath}
\usepackage{amssymb}
\usepackage{epsfig}
\usepackage{pspicture}
\usepackage{graphicx}
\usepackage{graphics}
\usepackage{booktabs}
 \usepackage{pst-grad} 
 \usepackage{pst-plot} 

\begin{document}

\title{A Three-Dimensional Model of Small Signal Free-Electron Lasers}

\author{Stephen Webb}
\email{swebb@grad.physics.sunysb.edu}
\affiliation{Department of Physics \& Astronomy, Stony Brook University}
\altaffiliation{Collider-Accelerator Department, Brookhaven National Laboratory}

\author{Gang Wang}
\email{gawang@bnl.gov}
\affiliation{Collider-Accelerator Department, Brookhaven National Laboratory}

\author{Vladimir Litvinenko}
\email{vl@bnl.gov}
\affiliation{Collider-Accelerator Department, Brookhaven National Laboratory}

\date{\today}

\begin{abstract}
Coherent electron cooling is an ultra-high-bandwidth form of stochastic cooling which utilizes the charge perturbation from Debye screening as a seed for a free-electron laser. The amplified and frequency-modulated signal that results from the free-electron laser process is then used to give an energy-dependent kick on the hadrons in a bunch. In this paper, we present a theoretical description of a high-gain free-electron laser with applications to a complete theoretical description of coherent electron cooling.
\end{abstract}

\maketitle

\section{Introduction}

Coherent electron cooling (CeC) \cite{litderb} is a new cooling method for intense relativistic hadron beams, to be implemented first at the proposed MEeRHIC/eRHIC upgrade to the RHIC accelerator at Brookhaven National Lab. Schematically similar to the stochastic cooling already implemented at RHIC \cite{blasbren}, CeC has the advantage that its coherent bandwidth is on the order of the resonance wavelength of the operating free-electron laser, so that the cross-correlation that leads to heating and therefore saturation of the stochastic cooling system is not encountered in CeC.

To achieve a complete theoretical description of Coherent Electron Cooling, models for the propagation of a phase space perturbation through the pick-up \cite{wangblas} and kicker \cite{gwang} were developed and presented \cite{gwang2}. All these calculations are based upon an infinite electron beam with $\kappa-2$ energy spread\footnote{A $\kappa$ distribution in a variable $x$ is a normalized distribution of the form $f_\kappa (x) \propto (1 + x^2/q^2)^{-\kappa}$ where $q$ is the spread parameter}. However, an exact analytical solution for the high gain free-electron laser in the small signal regime, given an initial phase space perturbation, had not yet been developed.

A number of analytical models have been developed for the transverse laser profile for an FEL. A set of equations for the full dynamics of a three-dimensional FEL with betatron oscillations were first written down in \cite{kjkim2}. Universal scaling for the gain of the FEL in terms of the energy spread, emittance and focusing properties were developed in \cite{ykg}. A fully three-dimensional Maxwell-Vlasov equation was studied in \cite{ckx} and ultimately a procedure for exact and variational solutions to the laser eigenmodes was presented in \cite{xie1}. These results focus primarily upon an eigenmode of the generated laser field, without consideration for developing solutions to the phase space density of the electron bunch. In this paper, we present a theoretical picture of the full dynamics of the electron phase space distribution, neglecting betatron oscillations, with an intent of using this result in application to CeC.

In Section 2, we present an overview of the configuration of Coherent Electron Cooling, and discuss briefly the existing results in the pick-up and kicker sections. With the context of this work in mind, we then present a derivation for the dynamics of a high-gain free-electron laser seeded with an initial phase space perturbation in Section 3. This leads to an equation for an arbitrary transverse distribution of an otherwise infinitely long electron beam. In Section 4, we analyze the case of an infinitely wide beam, which leads to a Green function for the 3D FEL process with an infinitely wide beam. A mode expansion method is considered for a finite beam in Section 5. To conclude, we consider the specifics of applying these results to Coherent Electron Cooling in Section 6.

\section{Overview of Coherent Electron Cooling}

Coherent Electron Cooling is schematically identical to stochastic cooling \cite{vdmeer80}, with a pick-up which gathers information about the position and energy of the individual particles in the hadron beam, an amplifier which takes this signal and amplifies it, and then a kicker which takes this information and uses it to deliver an energy-dependent non-conservative kick which decreases the longitudinal energy spread of the hadron beam.

\begin{figure}[htbp]
\begin{center}
\includegraphics*[width=90mm]{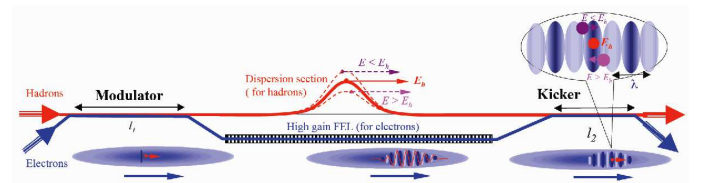}
\caption{Schematic of Coherent Electron Cooling}
\label{cecfig}
\end{center}
\end{figure}

For CeC, the pick-up is a co-moving electron bunch and hadron beam in a drift, where the individual hadron signals are the Debye screened charge perturbations described in \cite{wangblas}. The amplifier of the signal is the free-electron laser which we describe in this paper. The kicker is a chicane which offsets the hadrons from their initial signals so that they are displaced from a local maximum of the electron density in such a way that hadrons with energy greater than the design energy lose energy, whereas hadrons with energy less than the design energy gain energy. In both the kicker and the pick-up, it is necessary that the time co-moving between the hadrons and the electrons be shorter than a full plasma oscillation or else the signal will be greatly diminished.

Because the bandwidth of the coherent kicks from the amplified signal is on the order of the resonant wavelength of the FEL, which for most CeC applications is on the order of a few hundred nanometers, the cross-coherence that arises in stochastic cooling is negligible. Thus, the cooling system will continue to reduce the energy spread of the hadron beam until another effect is encountered.

To describe the FEL process, we present a theory that follows closely the derivation for the FEL instability derived in \cite{ssy}, with the slight modification that the transverse profile of the electron beam is uniform. We then inject this result into the existing results for the kicker and pick-up, and determine an exact form for the cooling decrement. But first we begin with a single-particle description of the dynamics.

\section{Maxwell-Vlasov Equations}

Consistent with the derivations of the high gain FEL in \cite{bpn} and \cite{ssynima} and summarized in \cite{ssy}, we begin with the equations of motion for the single particles in an undulator subject to the radiation field generated by the collective dynamics of the rest of the beam. The hamiltonian equations of motion for small energy deviation and high energy ($\gamma \gg 1$) are given by
\begin{subequations}
\begin{equation}\label{hequation}
\frac{d \mathcal{H}}{d z} = \frac{1}{c} \left \{ - \frac{1}{p_0} \left ( \frac{e}{c} \right)^2 \vec{A}_w \cdot \vec{A}_\perp + \frac{e}{c} \frac{\partial A_z}{\partial t} \right \}
\end{equation}
\begin{equation}\label{tequation}
\frac{d t}{d z} = \frac{1}{c} \left \{ 1 + \frac{1}{2} \frac{1}{p_0^2} \left [ \left ( \frac{e}{c} \right )^2 (\vec{A}_w^2 + 2 \vec{A}_w \cdot \vec{A}_\perp ) + m^2 c^2 \right ] \right \}
\end{equation}
\end{subequations}
where $p_0 = \mathcal{H}/c$, $\vec{A}_w = B_w /k_w (\cos k_w z ~ \hat{e}_y - \sin k_w z ~ \hat{e}_x)$ is the undulator vector potential (here we only consider helical undulators), $\vec{A}_\perp$ is the laser field and $A_z$ is the longitudinal space charge. The scalar potential has been removed by choice of gauge transformation, and $p_z$ has been used as the generator of longitudinal translations.

The Vlasov equation is derived from the conservation of single particle phase space volume, so that
\begin{equation}
\frac{d f}{d z} = \frac{\partial f}{\partial z} + \frac{d \mathcal{H}}{d z} \frac{\partial f}{\partial \mathcal{H}} + \frac{d t}{d z} \frac{\partial f}{\partial t} = 0
\end{equation}
Following along with the canonical description of instabilities in plasmas \cite{vlas67}, we assume that the phase space density of the electron beam is given by $f = f_1 + f_0$, where $f_0$ is a thermal background and $f_1$ is the instability. Furthermore, we assume that $|f_1| \ll |f_0|$. This justifies (i) dropping the term proportional to $\vec{A}_\perp^2$ that would appear in equation (\ref{tequation}) and (ii) dropping terms proportional to $f_1^2$ or higher. Carrying these approximations out and knowing that $\vec{A}_\perp, A_z \propto f_1$, we obtain the equation of motion given by
\begin{equation}
\begin{split}
\frac{\partial f_1}{\partial z} + \frac{1}{c} \left \{ 1 + \frac{1}{2} \frac{1}{\gamma_0^2} \left (1 + K^2 \right ) \left (1 - 2 \frac{\mathcal{E}}{\mathcal{E}_0} \right ) \right \} \frac{\partial f_1}{\partial t} \\ +  \left \{  \frac{1}{\mathcal{E}_0} \left ( \frac{e}{c} \right )^2 \vec{A}_w \cdot \frac{\partial \vec{A}}{\partial t} + e E_z \right \} \frac{\partial f_0}{\partial \mathcal{E}} = 0
\end{split}
\end{equation}
where $K = e A_w/m_e c^2$ is the undulator parameter. Absent from this description is an accounting for the transverse betatron oscillations that arise from the confining FODO lattice used on the electron beam in the undulator. In fact, all the transverse dynamics of this theory arise from the Maxwell equations, and it is assumed that the current distribution will follow this transverse distribution.

By solving the single-particle equations of motion for an electron in an undulator, this leads to the relationship
\begin{equation}
\vec{j}_\perp = \frac{K}{\gamma_0}\left ( \begin{array}{c} \cos k_w z \\ \sin k_w z \end{array}    \right ) j_z
\end{equation}
where $j_z \approx - e c \int d \mathcal{H} f_1(\mathcal{H}, z, t)$ is the longitudinal current density. We consider the transverse laser field in Fourier space, where its Fourier transform is defined by
\begin{equation}
\vec{A}_\perp = \frac{1}{\sqrt{2 \pi}^3} \int d\nu ~ d^2 k_\perp ~ e^{\imath \vec{k}_\perp \cdot \vec{r}_\perp} e^{\imath \nu \omega_r (z/c - t)} \tilde{A}_\perp (z, \nu, \vec{k}_\perp)
\end{equation}
The transverse Maxwell equation, when Fourier transformed over $\vec{r}_\perp$, is given by
\begin{equation}
\begin{split}
\frac{1}{\sqrt{2 \pi}^3}\int d \nu d^2 k_\perp \left ( - k_\perp^2 + \partial_z^2 - \frac{1}{c^2} \partial_t^2 \right ) \times \\ \left ( \tilde{A}_\perp e^{\imath \nu \omega_r (z/c - t)} e^{\imath \vec{k}_\perp \cdot \vec{r}_\perp} \right ) = \frac{4 \pi}{c} \vec{j}_\perp
\end{split}
\end{equation}
It is assumed that the envelope function $\tilde{A}_\perp$ is slow-varying in the longitudinal direction, and so higher order derivatives in $z$ are small compared to the first derivative. This allows us to drop terms that go as $\partial_z^2 \tilde{A}_\perp$ over $k_r \partial_z \tilde{A}_\perp$.

By dropping oscillating terms that are $2 k_w z$ out of phase with the laser field and defining the Fourier transform on $j_z$ by
\begin{equation}
\begin{split}
j_z = \frac{1}{\sqrt{2 \pi}^{3}} \int d\nu ~ d^2 k_\perp e^{\imath \vec{k}_\perp \cdot  \vec{r}_\perp} e^{\imath \nu \omega_r (z/c - t)}\times \\ e^{\imath k_u z}  e^{ - \imath k_\perp^2 c z/2 \nu \omega_r} \tilde{j}_z
\end{split}
\end{equation}
we obtain for $\vec{A}_w \cdot \vec{A}_\perp$ in Fourier space the expression
\begin{equation}
\vec{A}_w \cdot \tilde{A}_\perp = - e^{- \imath k_\perp^2 c z/2 \nu \omega_r} e^{\imath k_w z}\frac{\imath \pi K}{\nu \omega_r \gamma_0} \int_0^z \tilde{j}_z dz'
\end{equation}
where the initial laser field has been set to zero as is the case for CeC.

For the proof of principle, space charge will be a non-negligible component of the system. To account for space charge, we consider the longitudinal electric field given by
\begin{equation}
\partial_t E_z = - \frac{4 \pi}{c} j_z
\end{equation}
which, under this Fourier transform, gives
\begin{equation}
\tilde{E}_z = - \frac{4 \pi \imath}{c \nu \omega_r} \tilde{j}_z
\end{equation}
All of this is identical to the one-dimensional theory in \cite{ssy} except the additional phase factor of $k_\perp^2 c z/ 2 \nu \omega_r$ that appears in the definition of $\tilde{j}_z$, which acts as a detuning.

By applying an identical Fourier transform of the type performed on the current density to the phase space density, and assuming that the thermal background is given by
\[f_0 = n_0 F(\mathcal{E}) R(\vec{r}_\perp) \]
 we obtain the coupled Maxwell-Vlasov equation for the phase space density of the FEL amplified electron bunch with an initial phase space perturbation:
 \begin{widetext}
\begin{equation}\label{phasespaceeqn}
\begin{split}
\tilde{f}_1(\hat{z}, \nu, \vec{k}_\perp) = e^{-\imath \left ( k_w (1-\nu) + 2 k_w \nu \mathcal{E}/\mathcal{E}_0 - k_\perp^2 c/2 \nu \omega_r \right ) z} \tilde{f}_1 \mid_0 +  \int_0^z dz'~ e^{\imath \left ( k_w (1-\nu) + 2 k_w \nu \mathcal{E}/\mathcal{E}_0 - k_\perp^2 c/2 \nu \omega_r \right )( z'-z)} \times \\ \int d^2 q ~ e^{\imath \frac{(k_\perp^2 - q^2) c}{2 \nu \omega_r} z'} 
\left \{ \frac{\imath \nu \omega_r}{\mathcal{E}_0 c} \frac{e^2}{c^2} \left (- \frac{\imath \pi K}{\nu \omega_r \gamma_0} \int_0^{z'} dz''~ \tilde{j}_z (\vec{q}) \right ) - e \frac{4 \pi \imath}{c \nu \omega_r} \tilde{j}_z (\vec{q} )\right \} n_0 \frac{d F}{d\mathcal{E}} \tilde{R}(\vec{q} - \vec{k}_\perp )
\end{split}
\end{equation}
\end{widetext}
where $F = F(\mathcal{E})$ is the normalized energy distribution and $\mathcal{H} = \mathcal{E} + \mathcal{E}_0$ where $\mathcal{E}_0$ is the average energy of the electron beam. $\tilde{R}$ is the Fourier transform of the transverse bunch profile.

The equation of motion is identical in form to that of the one-dimensional theory in \cite{ssy}, with the exception of the added transverse detuning term $k_\perp^2 c / 2 \nu \omega_r$. Regardless of whether the beam is infinite or finite in transverse extent, the inverse gain length, given by \[ \Gamma =  \left ( \frac{ \mathcal{E}_0^2 c^2 \gamma_0}{2 \pi \nu e^3 K k_w n_0} \right )^{-1/3} \] and Pierce parameter, given by \[ \rho = \Gamma k_w^{-1}\] are unaffected by the three-dimensional effects.

To obtain the longitudinal current density, we take the definition $\tilde{j}_z \approx -e c \int d \mathcal{E} \tilde{f}_1$ to equation (\ref{phasespaceeqn}). Introducing the normalized detuning, space charge parameter, energy and transverse wave vector as
\begin{subequations}
\begin{equation}
\hat{C} = (1- \nu)/\rho
\end{equation}
\begin{equation}
\hat{\Lambda}_p^2 = \frac{8 \pi e^2 n_0 \Gamma^{-1} (1 + K^2)}{\gamma_0^3 m c^3} 
\end{equation}
\begin{equation}
\hat{\mathcal{E}} = 2 \nu \mathcal{E}/ \rho \mathcal{E}_0
\end{equation}
\begin{equation}
\hat{k}^2 = k^2 c \Gamma^{-1}/2 \nu \omega_r
\end{equation}
\end{subequations}
gives the cleaner and dimensionless form
\begin{equation} \label{currenteqn}
\begin{split}
\tilde{j}_z (\hat{z}, \hat{C}, \vec{k}_\perp) = - e c \frac{\rho \mathcal{E}_0}{2 \nu} \int d \hat{\mathcal{E}} ~ e^{\imath (\hat{C} + \hat{\mathcal{E}} - \hat{k}_\perp^2 ) \hat{z}} \tilde{f}_1 \mid_{\hat{z} = 0} + \\
\int d\hat{\mathcal{E}} ~ \int_0^{\hat{z}} d\hat{z}' e^{\imath (\hat{C} + \hat{\mathcal{E}} - \hat{k}_\perp^2) (\hat{z}' - \hat{z})} \int d^2 \hat{q} ~ e^{\imath (\hat{q}^2 - \hat{k}_\perp^2) \hat{z}'} \times \\ \left \{ \int_0^{\hat{z}'} d\hat{z}'' ~ \tilde{j}_z (\vec{q}) + \imath \hat{\Lambda}_p^2 \tilde{j}_z (\vec{q}) \right \} \frac{d \hat{F}}{d \hat{\mathcal{E}}} \hat{R}(\vec{q} - \vec{k}_\perp)
\end{split}
\end{equation}

At this point, the method of solution depends on whether the beam is to be considered finite or infinite in transverse size, which is to say whether the transverse dimension of the electron bunch $r_0$ is large compared to the diffraction length scale of the FEL, $d = \sqrt{ c \Gamma^{-1}/2 \nu \omega_r}$.

On the other hand, whether the transverse spacial extent of the initial perturbation can be modeled profitably as a delta function in real space (which would be much simpler) depends on the comparison of the Debye radius to the transverse length scale, $r_D/d$. If the Debye radius is much smaller than $d$, $r_D/d \ll 1$, then the physics of a point-perturbation in transverse space should match very closely the physics of the initial phase space perturbation. If $r_D/d \sim 1$ then the actual physical distribution is necessary. If $r_D/d \gg 1$ then we expect the FEL to be essentially one-dimensional. By necessity, $r_D \ll r_0$ for the models utilized in \cite{wangblas} and \cite{gwang2} to be valid. These considerations hold for both the infinite and finite beam solutions.

\section{Infinite Beam Size}

We first consider a beam that is infinite in the transverse direction, as it is analytically simpler than the finite beam size but still contains a reasonable amount of physics in its own right. This can be considered in terms of the ratio $r_0/d$, where $r_0$ is the typical transverse width scale of the electron beam and $d$ is the diffraction length scale of the FEL. If $r_0/d \gg 1$ then the beam is effectively infinite and the treatment in this section is useful. Otherwise the finite beam solution of the next section needs to be employed.

For an infinite beam, $\tilde{R}(\vec{q} - \vec{k}_\perp) = \delta (\vec{q} - \vec{k}_\perp)$ so the above equation (\ref{phasespaceeqn}) reduces to
\begin{equation}\label{currenteqn}
\begin{split}
\tilde{j}_z = - e c \frac{\rho \mathcal{E}_0}{2 \nu} \int d \hat{\mathcal{E}} ~ e^{\imath (\hat{C} + \hat{\mathcal{E}} - \hat{k}_\perp^2 ) \hat{z}} \tilde{f}_1 \mid_{\hat{z} = 0} + \\
\int d\hat{\mathcal{E}} ~ \int_0^{\hat{z}} d\hat{z}' e^{\imath (\hat{C} + \hat{\mathcal{E}} - \hat{k}_\perp^2) (\hat{z}' - \hat{z})}  \left \{ \int_0^{\hat{z}'} d\hat{z}'' ~ \tilde{j}_z + \imath \hat{\Lambda}_p^2 \tilde{j}_z \right \} \frac{d \hat{F}}{d \hat{\mathcal{E}}}
\end{split}
\end{equation}

This is identical in form to the equations of motion for the one-dimensional FEL \cite{ssy} with the identification of $\hat{C}_{3D} = \hat{C} - \hat{k}_\perp^2$. Due to this similarity, we omit many of the details and cut to the solution by Laplace transform for the current, which is given by
\begin{equation}\label{currentlaplace}
\mathcal{J}(s) =  \frac{- e c \frac{\rho \mathcal{E}_0}{2 \nu} \int d \hat{\mathcal{E}} ~\frac{1}{s + \imath (\hat{C}_{3D} + \hat{\mathcal{E}}) }\tilde{f}_1 \mid_{\hat{z} = 0} }{s - \hat{D} (1 - \imath s \hat{\Lambda}_p^2)}
\end{equation}
where
\begin{equation}
\hat{D} = \int d \hat{\mathcal{E}} ~ \frac{d \hat{F}}{d \hat{\mathcal{E}}} \frac{1}{s + \imath (\hat{C}_{3D} + \hat{\mathcal{E}})}
\end{equation}
determines the dispersion relation. Equation (\ref{currentlaplace}) gives immediately the linear response function in Laplace space for the current density perturbation versus an initial phase space perturbation
\begin{equation}
\tilde{j}_z = \int d \hat{\mathcal{E}}' \mathcal{K} (s, \hat{C}_{3D}, \mathcal{E}')\tilde{f}_1 \mid_0 (\mathcal{E}')
\end{equation}
such that
\begin{equation}
\mathcal{K} (s, \hat{C}_{3D}, \mathcal{E}') = - e c \frac{\rho \mathcal{E}_0}{2 \nu} \frac{1}{s - \hat{D}(1 - \imath s \hat{\Lambda}_p^2)} \frac{1}{s + \imath(\hat{C}_{3D} + \mathcal{E}')}
\end{equation}
$\mathcal{K}$ is the linear response function of the modulated current density to an initial phase space perturbation. We will use this function to calculate a Green function for the FEL phase space distribution, which we will denote $\mathcal{G}_{FEL}$.

By inserting equation (\ref{currentlaplace}) back into equation (\ref{phasespaceeqn}), and Laplace transforming for $\tilde{f}_1$ in the $\hat{z}$ coordinate, we obtain a comparable expression to eqn. (\ref{currentlaplace}) for the phase space density of the perturbation of the e-beam given by:
\begin{equation}
\begin{split}
\tilde{f}_1(s, \hat{C}_{3D}, \hat{k}_\perp, \hat{\mathcal{E}}) = \frac{1}{s + \imath (\hat{C}_{3D} + \hat{\mathcal{E}})} \tilde{f}_1 \mid_0 (\mathcal{E}, \hat{C}_{3D}, \hat{k}_\perp) + \\ \frac{1}{s + \imath (\hat{C}_{3D} + \hat{\mathcal{E}})} \left [ \frac{1}{s - \hat{D} (1 - \imath s \hat{\Lambda}_p^2 )} + \imath \hat{\Lambda}_p^2 s \frac{1}{s - \hat{D} (1 - \imath s \hat{\Lambda}_p^2)} \right ] \\ \frac{d \hat{F}}{d \hat{\mathcal{E}}} \int d\hat{\mathcal{E}}' \tilde{f}_1 \mid_0 (\hat{\mathcal{E}'}, \hat{C}_{3D}, \hat{k}_\perp) \frac{1}{s + \imath (\hat{C}_{3D} + \hat{\mathcal{E}}')} 
\end{split}
\end{equation}

The form of this equation allows us to write down the Green function for the phase space density of an infinitely wide e-beam in an FEL amplifier as
\begin{widetext}
\begin{equation}\label{greenfunction}
\begin{split}
\mathcal{G}_{FEL} (s, \hat{C}_{3D}, \hat{k}_\perp, \hat{\mathcal{E}}; \hat{\mathcal{E}}') = \frac{1}{s + \imath (\hat{C}_{3D} + \hat{\mathcal{E}})} \delta(\hat{\mathcal{E}} - \hat{\mathcal{E}}') + \dots \\ \frac{1}{s + \imath (\hat{C}_{3D} + \hat{\mathcal{E}})} \left [ \frac{1}{s - \hat{D} (1 - \imath s \hat{\Lambda}_p^2 )} + \imath \hat{\Lambda}_p^2 s \frac{1}{s - \hat{D} (1 - \imath s \hat{\Lambda}_p^2)} \right ] \frac{d \hat{F}}{d \hat{\mathcal{E}}} \frac{1}{s + \imath (\hat{C}_{3D} + \hat{\mathcal{E}}')} 
\end{split}
\end{equation}
\end{widetext}
where the new FEL phase space density in Laplace-Fourier space is given by
\begin{equation}
\begin{split}
\tilde{f}_1 (s, \hat{C}_{3D}, \hat{k}_\perp, \hat{\mathcal{E}}) = \\ \int d \hat{\mathcal{E}}'~ \mathcal{G}_{FEL}(s, \hat{C}_{3D}, \hat{k}_\perp, \hat{\mathcal{E}} | \hat{\mathcal{E}}' ) \tilde{f}_1 (\hat{C}_{3D}, \hat{k}_\perp, \hat{\mathcal{E}}') \mid_0
\end{split}
\end{equation}
It is interesting to note that this Green function can be clearly divided into two parts. The first part represents Landau damping and single-particle non-cooperative motion in the FEL undulator. This process does not lead to gain, and the term representing it can be dropped in a description of the FEL process. The second part contains the growing roots of the dispersion relation, and represents the cooperative gain process of the FEL. It is this Green function that is of practical application for the theory of Coherent Electron Cooling.

The dynamics in the $\hat{z}$ variable are determined by the roots of the dispersion relation, given by
\begin{equation} \label{dispersion}
s - \frac{\hat{D}}{1 - \imath \hat{\Lambda}_p^2 \hat{D}} = 0
\end{equation}
There is another pole from the $s + \imath (\hat{C}_{3D} + \mathcal{E})$ term in the denominator, but the pole associated with this term will either oscillate or decay, and therefore does not represent amplification as a result of the FEL process, but rather a Landau damping of the initial perturbation due to its own energy spread.

As an example calculation, we consider an initial phase space perturbation that is monoenergetic, instantaneous in time, and a point source. We place this in the context of a cold electron beam, where the dispersion relation is well known.

In Fourier space, the transform of the initial condition is given by
\begin{equation}
\tilde{f}_1 \mid_0 = \delta(\hat{\mathcal{E}} - \hat{\mathcal{E}}_0)
\end{equation}
where it is infinitely broad in the $\vec{k}_\perp$ and $\hat{C}$ variables.

Inserting this directly into the Green function calculation and taking the inverse Laplace transform on $s$ gives a sum with three purely oscillating terms and with the three modes of the FEL process. The resulting expression is extremely cumbersome, and its physical intuition is embodied already in the Green function. We therefore only consider the single growing root of the FEL process from here on.

The phase space density is then approximately given by
\begin{equation}
\begin{split}
\tilde{f}_1 (\hat{z}, \hat{C}_{3D}) \approx \frac{1}{s_+ + \imath (\hat{C}_{3D} + \hat{\mathcal{E}})} \frac{1}{s_+ + \imath (\hat{C}_{3D} + \hat{\mathcal{E}}_0)} \\ \left \{ \frac{1 + \imath \hat{\Lambda}_p^2 s_+}{1 - \hat{D}'\mid_{s+} (1 - \imath s_+ \hat{\Lambda}_p^2) - \hat{D}\mid_{s_+} } \right \} \frac{d \hat{F}}{d \hat{\mathcal{E}} } \exp \left (s_+ \hat{z} \right )
\end{split}
\end{equation}
where $s_+$ is the root of the dispersion relation with positive real value. Expectedly, all dependence on $\vec{k}_\perp$ has dropped out, and only $\hat{C}_{3D}$ remains as the natural Fourier parameter for the infinite electron beam.

Recall the definition of the Fourier transformed phase space density as:
\begin{equation}
\begin{split}
f_1 (z, z/c - t, \vec{r}_\perp, \mathcal{E}) = \\ \frac{1}{\sqrt{2 \pi}^{3}} \int d\nu ~ d^2 k_\perp e^{\imath \vec{k}_\perp \cdot  \vec{r}_\perp} \times \\e^{\imath \nu \omega_r (z/c - t)} e^{\imath k_u z}  e^{ - \imath k_\perp^2 c z/2 \nu \omega_r} \tilde{f}_1 (z, \nu, \vec{k}_\perp, \mathcal{E})
\end{split}
\end{equation}
It would now be useful to transform the integrals into integrals over $\hat{C}_{3D}$ and $\hat{k}_\perp$ in order that we can determine the dynamics of this initial perturbation in real space. Recalling the definitions of the parameters leaves
\begin{equation}
\begin{split}
f_1(\hat{z}, \xi, \hat{r}_\perp, \hat{\mathcal{E}}) = - \frac{1}{\sqrt{2 \pi}^{3}} \frac{2 \omega_r}{c \rho \Gamma^{-1}} e^{\imath \xi} 
\int d\hat{C}_{3D} ~ d^2 \hat{k}_\perp \\ e^{\imath \hat{k}_\perp \cdot \hat{r}_\perp} 
e^{-\imath \rho (\hat{C}_{3D} - \hat{k}_\perp^2) \xi} e^{- \imath \hat{k}_\perp^2 \hat{z}} \tilde{f}_1 (\hat{C}_{3D}, \hat{z}, \hat{\mathcal{E}}) + c.c.
\end{split}
\end{equation}
where $\xi = \omega_r(z/c - t) + k_u z$ is the ponderomotive phase. It is interesting to note that, although the detailed temporal information cannot be extracted from this integral immediately, the transverse profile can be calculated directly as
\begin{equation}
\tilde{f}_1 \propto \frac{1}{\hat{z} + \rho \xi}  e^{- \imath r_\perp^2/(4(\rho \xi + \hat{z})} \end{equation}
Because this is a pure phase, it has no transverse size information intrinsic to it. The trouble arises from the equal value given to all $\vec{k}_\perp$ by an infinitely small point source, which allows the signal to propagate transversely instantly as we have not properly accounted for a Lorentz covariant description of the transverse electron dynamics.

A slightly less mathematically pathological case is to consider an initially Gaussian transverse distribution, infinitely short. In this case, the previous separation also occurs and the resulting width goes as $\sigma_{\hat{r}}^2 \sim \hat{\sigma}_0^2 - \imath (\rho \xi + \hat{z})$ and the profile is gaussian rather than sinusoidal.

\section{Finite Beam Size}

Having considered the simpler case of the transversely infinite beam, we now turn our attention to the case of a finite transverse beam profile. To achieve this, we consider an expansion in the eigenmodes of the transverse beam profile, as the Maxwell-Vlasov equation for a finite beam is an integral equation with the beam profile function as its kernel. From there, we can separate out the transverse and longitudinal dynamics, and observe that in real space there is no spreading of the eigenmodes, consistent with optical guiding.

\subsection{Eigenmode Expansion}

For the case when $\tilde{R}$ is not a delta function, it is beneficial to expand the current density solutions in the eigenmodes of the $\tilde{R}$ kernel defined by
\begin{equation}\label{eigeneqn}
\psi_\ell(\vec{k}) = \frac{1}{\omega_\ell} \int d^2 \vec{q} ~ \tilde{R}(\vec{k} - \vec{q}) \psi_\ell (\vec{q})
\end{equation}
where for this section we drop the overhats and subscripts to simplify the notation. This is best calculated by expanding $\tilde{R}(\vec{k} - \vec{q})$ as a matrix in terms of some orthonormal basis. We shall consider such an example calculation later, but for now we assume such an eigenbasis is already known.

For any reasonably smooth definition of the transverse beam profile, $\tilde{R}(\vec{k} - \vec{q}) = \tilde{R}(\vec{q} - \vec{k})$, that is that the kernel of the eigenvalue equation is hermitian \cite{tricomi}. This being the case, we know that the eigenvectors are orthogonal and the eigenvalues are all real, with the orthogonality condition being
\begin{equation}
\int d^2q d^2 k ~ \psi_{\ell}(\vec{k}) \psi_{m}(\vec{q}) = \delta_{\ell m}
\end{equation}
given that the eigenfunctions are square normalized.

Expanding the integral of the longitudinal current density in a series of the eigenmodes gives
\begin{equation}
\int^{\hat{z}}_0 d\hat{z}' ~ \tilde{j}_z (\hat{z}') = \sum_\ell \psi_\ell (\vec{k})_\perp e^{\imath k^2 \hat{z}} a_\ell (\hat{z})
\end{equation}

Looking back at the definition of the Fourier transform for the current, it is clear that the $e^{\imath k_\perp^2 \hat{z}}$ terms will cancel, and there is no change in the transverse extent of the current perturbation, which is consistent with the optical guiding discussed in the literature \cite{scharlemann}.

The current equation (\ref{currenteqn}) can then be reduced to a system of coupled equations for the expansion coefficients. That equation is given by
\begin{equation}
\begin{split}
 a_\ell' - \imath Q_{m, \ell} a_m = \\ - e c \frac{\rho \mathcal{E}_0}{2 \nu} \int d \hat{\mathcal{E}} \int d^2 \hat{k}_\perp ~ e^{\imath (\hat{C} + \hat{\mathcal{E}} - \hat{k}_\perp^2) \hat{z}} \tilde{f}_1\mid_0 \psi_\ell(\hat{k}_\perp) - \\
 \int d\hat{\mathcal{E}} \int_0^{\hat{z}} d\hat{z}' ~ e^{\imath (\hat{C} + \hat{\mathcal{E}})(\hat{z}' - \hat{z})} \times \\
\frac{1}{\omega_\ell} \biggl \{ a_n  
+ \imath \hat{\Lambda}_p^2 \left [ a_\ell' + \imath Q_{m, \ell} a_m \right ] \biggr \} \frac{d \hat{F}}{d \hat{\mathcal{E}}}
\end{split}
\end{equation}
where $Q_{m, \ell} = \int d^2 k ~ k^2 \psi_m (\vec{k}) \psi_\ell (\vec{k})$ measures the coupling between the different eigenmodes of the transverse beam profile. This system of equations may be solved by Laplace transform (see the Appendix for details) and leads to the equation
\begin{equation}\label{finitecoeff}
\begin{split}
\biggl [\left ( s - \hat{D} \omega_m (1 + \imath s \hat{\Lambda}_p^2 ) \right ) \delta_{\ell, m} + \dots \\ (1 + \imath \hat{\Lambda}_p^2 \omega_m ) Q_{\ell, m} \biggr ] a_m =
 \tilde{f}_1^\ell 
\end{split}
\end{equation}
where $\tilde{f}_1^\ell$ is the $\psi_\ell$ component of the initial phase space perturbation, and
\begin{equation}\label{dispersion}
\hat{D} = \int d \hat{\mathcal{E}} ~ \frac{d \hat{F}}{d \hat{\mathcal{E}}} \frac{1}{s + \imath (\hat{C} + \hat{\mathcal{E}})}
\end{equation}
determines the dispersion relation for each growing mode.

It is worth noting, at this point, that the $\ell$ index could refer to multiple indices, particularly since this is a two-dimensional model it could refer to both the azimuthal and axial indices, as will be the case when we consider the Gaussian beam profile below. For that particular case, the different azimuthal modes are uncoupled in the $Q$ matrix, so the radial modes for a particular azimuthal mode are the ones coupled by the $Q$ matrix, while differing azimuthal modes do not mix. This will become apparent during the calculation below.

\subsection{Gaussian Profile}
As an example of this calculation, we consider a Gaussian transverse beam profile. The procedure for solving the initial value problem is as follows:
\begin{enumerate}
\item Calculate the eigenfunctions and corresponding eigenvalues to equation (\ref{eigeneqn})
\item Calculate $Q_{m, \ell}$ to determine the correct dispersion relation
\item Invert equation (\ref{finitecoeff}) and solve for the initial value problem
\end{enumerate}
Each of these steps should be identical for any other transverse bunch profiles; we present only the Gaussian case here.

We begin with a Gaussian beam profile, whose Fourier transform is given by
\begin{equation}
\tilde{R}(\vec{k}_\perp - \vec{q}) = \left (\frac{\hat{L}}{\sqrt{2 \pi}} \right )^2\exp \left \{ - \frac{ (\vec{k}_\perp - \vec{q})^2}{2 \hat{L}^{-2}} \right \}
\end{equation}
and the eigenfunctions therefore satisfy the equation
\begin{equation}
\begin{split}
\psi_\ell (\vec{k}_\perp) = \\ \frac{1}{\omega_\ell} \int d^2 \vec{q} ~ \left (\frac{\hat{L}}{\sqrt{2 \pi}} \right)^2\exp \left \{ -\frac{(\vec{k}_\perp - \vec{q})^2}{2 \hat{L}^{-2}} \right \}  \psi_\ell (\vec{q})
\end{split}
\end{equation}
It is most convenient to consider this particular form in Cartesian coordinates, and in keeping with this we expand
\begin{equation}
\psi_\ell (\vec{p}) = \chi_m (p_x) \chi_n (p_y)
\end{equation}
where each of the individual $\chi$ satisfy an eigenvalue equation of the form
\begin{equation}
\begin{split}
\chi_m (p_\imath) = \frac{1}{\lambda_m} \int_{-\infty}^{\infty} dp'_\imath ~  \frac{\hat{L}}{\sqrt{2 \pi}} \\ \exp \left \{ - (p_\imath^2 + p_\imath'^2 - 2 p_\imath p'_\imath)/2 \hat{L}^{-2} \right \} \chi_m (p'_\imath)
\end{split}
\end{equation}
where the resulting eigenvalue for $\psi_\ell$ is given by $\omega_\ell = \lambda_n \lambda_m$. It is convenient to define the normalized variable $\mu = p_\imath \hat{L}$ so that the above eigenvalue equation is given by
\begin{equation}
\chi_m (\mu) = \frac{1}{\hat{\lambda}_m} \int_{-\infty}^{\infty} d\mu' ~ \\ \exp \left \{ - (\mu^2 + \mu'^2 - 2 \mu \mu')/2 \right \} \chi_m (\mu')
\end{equation}
where $\hat{\lambda}_m = \lambda_m \sqrt{2 \pi}$. The appropriate scaling for the transverse beam size for the full eigenvalue is given by \[\omega_\ell = \frac{\hat{\omega_\ell}}{2 \pi}\] where $\hat{\omega_\ell} = \hat{\lambda}_m \hat{\lambda}_n$. To calculate the normalized eigenvalues, we expand the kernel of this single-variable integral equation in terms of Hermite polynomials, as they are already related to the paraxial Maxwell equations \cite{nbc}.

It turns out from the properties of Hermite polynomials that only the evens and odds couple, so each $\chi_m$ is a series in either even or odd Hermite polynomials. In this case, the matrix equation for the even Hermite polynomials is given approximately by the matrix elements
\begin{equation}
\begin{split}
\mathtt{G}_{a, b} = \int_{-\infty}^{\infty} d\mu  \int_{-\infty}^{\infty} d\mu' ~  \exp \left \{ - (\mu^2 + \mu'^2 - 2 \mu \mu')/2 \right \}\\H_a(\mu) e^{-\mu^2/2} H_b(\mu') e^{-\mu'^2/2}
\end{split}
\end{equation}

Furthermore, to good approximation, the expansion can be carried out for the first two Hermite functions in the series. We therefore consider the two-mode case. For the principle even mode, the matrix is given by
\begin{equation}
\mathtt{G} = \left ( \begin{array}{cc} 2 \sqrt{\frac{\pi}{3}}& \frac{1}{3}\sqrt{ \frac{2 \pi}{3}} \\ \frac{1}{3}\sqrt{ \frac{2 \pi}{3}} & \sqrt{\frac{\pi}{3}} \end{array} \right )
\end{equation}
for the vector components $(H_0(\mu) , H_2(\mu) )^t \exp(-\mu^2/2)$. The eigensystem here has eigenvalue $\hat{\lambda}_{even} = 2.2382$ with corresponding eigenvector 
\[\vec{v}_{even} = \left ( \begin{array}{c} .9294 \\.3690 \end{array} \right ) \]
and a smaller eigenvalue $\hat{\lambda}_2 = .83178$ with corresponding eigenvector
\[ \left ( \begin{array}{c} -.1465 \\ .3690 \end{array} \right ) \]

To validate these numerical results we take the matrix to next order, \emph{i.e.} to order $H_4(\mu)$ in the expansion, and the matrix is given by
\begin{equation}
\mathtt{G} = \left ( \begin{array}{ccc}
 2 \sqrt{\frac{\pi}{3}}& \frac{1}{3}\sqrt{ \frac{2 \pi}{3}} & \frac{1}{9} \sqrt{\frac{\pi}{2}} \\
 \frac{1}{3}\sqrt{ \frac{2 \pi}{3}} & \sqrt{\frac{\pi}{3}} & \frac{17}{54} \sqrt{\pi} \\
 \frac{1}{9} \sqrt{\frac{\pi}{2}} & \frac{17}{54} \sqrt{\pi}  & \frac{227}{324} \sqrt{\frac{\pi}{3}}
  \end{array} \right )
\end{equation}
which yields an eigensystem given by $\hat{\lambda}'_1 = 2.3157$, $\hat{\lambda}'_2 = 1.2005$ and $\hat{\lambda}'_3 = .27073$ with corresponding normalized eigenvectors
\[ \vec{v}_1 = \left ( \begin{array}{c} .8772 \\.4244 \\.2245 \end{array} \right ) \]

\[ \vec{v}_2 = \left ( \begin{array}{c} -.1724 \\.2376 \\  .2245 \end{array} \right ) \]

\[ \vec{v}_3 = \left ( \begin{array}{c} .03343\\ - .1879 \\  .2245 \end{array} \right ) \]

We can conclude from this that the largest eigenvalue can be accurately determined to within $3 \%$ with the $2 \times 2$ matrix expansion, and from analysis of the eigenvector components the $H_4(\mu)$ level of expansion is negligibly small compared to the other two components for the eigenvector with the maximal eigenvalue.

Carrying out a similar procedure for the $H_1(\mu)$ - $H_3(\mu)$ eigenmode gives a maximal eigenvalue $\hat{\lambda}_{odd} = 1.7161$ and eigenvector
\[\vec{v}_{odd} = \left ( \begin{array}{c} .8456\\ .5339  \end{array} \right ) \]

\begin{figure}[htbp]
\begin{center}
\includegraphics*[width=90mm]{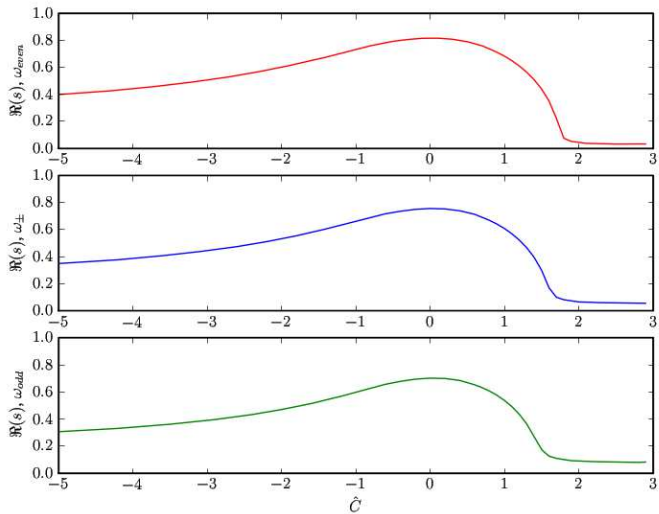}
\caption{Growth rates for three eigenmodes: (i) top is of mode with largest eigenvalue, (ii) is degenerate case of the odd/even mixtures, (iii) is of smallest eigenvalue}
\label{gainfig}
\end{center}
\end{figure}

It is now necessary to calculate the various matrix elements for $Q$. For the purposes orderly book-keeping, we define the following modes
\begin{subequations}
\begin{equation}
\psi_{even} = \chi_{even} (\mu_x) \chi_{even} (\mu_y)
\end{equation}
\begin{equation}
\psi_{odd} = \chi_{odd}(\mu_x) \chi_{odd}(\mu_y)
\end{equation}
\begin{equation}
\psi_+ = \frac{1}{\sqrt{2}} \left (\chi_{odd}(\mu_x) \chi_{even}(\mu_y) + \chi_{even}(\mu_x) \chi_{odd}(\mu_y) \right )
\end{equation}
\begin{equation}
\psi_- = \frac{1}{\sqrt{2}} \left (\chi_{odd}(\mu_x) \chi_{even}(\mu_y) - \chi_{even}(\mu_x) \chi_{odd}(\mu_y) \right )
\end{equation}
\end{subequations}
as the orthonormal basis of expansion. The corresponding eigenvalues are given by $\hat{\omega}_{even} = 5.0095$, $\hat{\omega}_{odd} = 2.945$ and $\hat{\omega}_{+} = \hat{\omega}_{-} = 3.8410$. Under this particular basis the Hermite polynomials have a particularly nice relation for the $Q$ matrix elements, and $Q$ is diagonal. The individual modes do not couple, and their growth rates are determined by the dispersion relation
\begin{equation}
\left (s - \hat{D} \omega_m (1 + \imath s \hat{\Lambda}_p^2 ) \right ) + (1 + \imath \hat{\Lambda}_p^2 \omega_m) Q_{m, m} = 0
\end{equation}
The individual $Q$ are given by $Q_{even} = 2.51446/\hat{L}^4$, $Q_{odd} = 6.35275/\hat{L}^4$, and $Q_{+} = Q_{-} = 4.43333/\hat{L}^4$. The growth rate for these parameters is given in figure (\ref{gainfig}), with $\hat{L} = 3$.

To recap, we have calculated an eigenbasis for the transverse beam profile, yielding a linear superposition of even- and odd-numbered Hermite polynomials, and their corresponding eigenvalues. The series is truncated at two dominant modes, and because of the particular nature of the Hermite polynomial expansion basis, the $Q$ matrix is diagonal. If $Q$ had off-diagonal matrix elements, there would be ``gain leakage'' between the connected eigenvectors.

\subsection{One-Dimensional Limit}

Because the eigenvalues are totally independent of the transverse size, and only $Q$ is dependent, it is straightforward to get directly to the one-dimensional beam limit for the dispersion relation. By redefining the normalization as
\begin{subequations}
\begin{equation} \tilde{s} = s \omega_m ^{-1/3} \end{equation}
\begin{equation} \tilde{C} = \hat{C} \omega_m ^{-1/3} \end{equation}
\begin{equation} \tilde{\Lambda}_p^2 = \hat{\Lambda}_p^2 \omega_m ^{1/3} \end{equation}
\begin{equation} \tilde{Q}_m = Q_m \omega_m ^{-1/3} \end{equation}
\end{subequations}
the dispersion relation takes the form
\begin{equation}
\tilde{s} - \frac{\imath}{(\tilde{s} + \imath \tilde{C})^2} (1 + \imath \tilde{s} \tilde{\Lambda}_p^2) + (1 + \imath \tilde{\Lambda}_p^2 \omega_m^{2/3}) \tilde{Q} = 0
\end{equation}
The actual scaling is such that, for large beams, the portion of this dispersion relation identical in form to the one-dimensional dispersion relation comes to strongly dominate over the perturbation correction for finite size, taken by the value of $Q_m$. For the case of an infinitely large transverse size all functions are eigenmodes and all all eigenvalues are unity, therefore we can obtain the one-dimensional limit through this limit.

\section{Discussion}

We have presented a theoretical model for the dynamics of a high-gain free-electron laser with three-dimensional effects. The model is analytically solvable up to a numerical Fourier transform, and for that reason is useful for benchmarking the massive tracking programs used to simulate FELs. All results in this paper are reduced to a handful of dimensionless numerical Fourier transforms.

When applying the finite beam case, we observe that only the principle four modes grow rapidly. The higher order modes have eigenvalues substantially smaller than these modes, and can be neglected in comparison to the principles. We can therefore conclude from this model that an FEL can be effectively characterized by only a handful of well-understood eigenmodes. Furthermore, this particular model includes optical guiding by consideration of the transverse eigenmodes of a stationary beam. By contrast, we observe spreading of the infinite beam case at a slower than linear rate.

The principle goal of this solution to the three-dimensional FEL equations is to develop an understanding of the charge modulation at the end of the undulator. A thorough understanding of the phase information of the FEL instability is necessary to properly calibrate the chicane and inject the hadrons with a proper displacement with respect to the local charge maxima of the bunch. This model provides the phase information up to a three-dimensional Fourier integral, which is well-bounded and provides adequate benchmarking for numerical simulations.

The existing analytical models for the kicker and pick-up of CeC involve an infinitely large electron beam, or equivalently that the initial perturbation be small compared to the transverse size of the electron beam. The results are also obtained analytically for the $\kappa-2$ distribution. To match up with these theories, we consider the case where $\tilde{R}(\hat{k}_\perp - \hat{q}) = \delta (\hat{k}_\perp - \hat{q})$ and with the corresponding dispersion relation for a $\kappa-2$ distribution. The results for $\kappa-2$ are not presented in this paper, but it is straightforward to obtain the dispersion relation from the dispersion integral, and we can now consider a complete description of the phase space evolution of the electron bunch through the CeC process.

This analytical model was developed to provide benchmarking for the proof of principle CeC system to be implemented at RHIC. For the FEL for the proof of principle, the transverse size of the electron bunch is $r_0 \approx 3 ~\textrm{mm}$, the resonant wavelength is $\lambda_r \approx .5 ~ \mu \textrm{m}$, and a gain length of approximately $\Gamma^{-1} = 3 ~ \textrm{m}$. In this case, the transverse length scale $d \approx .35 ~\textrm{mm}$ and it is expected that the three-dimensional infinite beam theory should be a reasonable description of the FEL amplifier portion of CeC.

At present this model has no way of coping with a transverse momentum spread in the initial phase space perturbation or with betatron oscillations, because all of the dynamics are taken directly from Maxwell's equations. As such, it is not clear what effect transverse momentum spread and betatron oscillations will have on the phase information of the amplified signal. Numerical modeling or a more complete theoretical description are necessary to account for these effects.

\section{Acknowledgements}

The authors would like to thank Michael Blaskiewicz and Evgeny Saldin for helpful discussion. Work supported by Brookhaven Science Associates, LLC under Contract No. DE-AC02-98CH10886 with the U.S. Department of Energy.


\bibliography{3dfel_cec_bib}

\end{document}